# Complex Coupled Mode theory (C-CMT) of planar multilayer waveguides

**SHAHRAM MORADI**[1, 2]

1Electrical and Computer Engineering, University of Victoria, Victoria, Canada
2Micro and Nano Technology Department, Middle East Technical University, Ankara, Turkey
Email: photonicsandoptics@gmail.com

**Abstract-**The numerical complex coupled-mode method used in a metal thin-film optic elements is applied to a planar multilayer optical waveguide. All modes are required to satisfy Helmholtz vectorial equation in an optical waveguide including bound and leaky modes at forward and backward propagation directions. First-Order perturbation theory is developed and applied by means of lossy materials including metal thin-film and perfectly matched layers at the boundaries. Our suggested structure discontinues in one direction and gives rise to an asymmetric metal thin-film at the core for investigating the transmission and reflection coefficient using mode-matching theory in two sides of the truncation. Finally, we compared the computed numerical results with the one that we obtained with the finite difference solver method. This study provides several advantages of CCMT in asymmetric temporal optical resonators that are in huge demand for today's integrated optical technology.

**Introduction**

One of the ubiquitous phenomenon in the field of photonics is resonance, in which resonant mode interacts with the propagated waves, the analyzing dynamics of the resonant system via coupled mode theory match with rigorous numerical simulations[1][2][3][4]. Coupled mode theory as an important widely used methods for analyzing coupling between discrete guided modes plays a major role in today's application. The analytic solutions in this method provides a reliable and applicable solutions in a broad range of applied physics such as tilted fiber Bragg gratings (TFBGs)[5][6], near-field thermophotovoltaics[7], acoustic resonators[8], non-Hermitian waveguides[9][10][11], quasi-single mode fibers [12], linear[13] and non-linear[14][6] electromagnetic resonators ad so many others. On the other hand, the complex coupled mode theory[15] [16] [17] as a full vector analysis tool[5] is a developed method that possesses salient characteristics which make it distinct from the conventional one. In terms of comparison, CMT is not a proper analytical methods for some structures such as step-index slab waveguide and circular fibers since radiation modes and/or considering leaky modes which include the continuum of the radiation fields under certain conditions[18]. Thus, CMT is not fit for such conditions as it does for discrete guided modes[19][20], and it is difficult to deal with lossy materials with the leaky modes since the continuum of radiation is not orthogonal and normalizable in real domain[21][22].

Here in this study, a full vector complex coupled mode theory[23][24] for a 2D waveguide is developed and presented. The discrete modes are decomposed by complex effective indexes of mode solved from mode solver with considering perfectly matched layer (PML) and the perfectly electric conductors (PEC) boundary. By considering the PML and PEC, an ideal waveguide model is established for which physical domain of the suggested structure is not influenced via the interference of the reflected waves from the boundaries of the computation window. We divide the computational domain to two sides with discontinuity of metal film at the core. Then we solve the modes for both sides of the suggested structure separately and using mode matching to investigate contribution of computed eigenmodes in the whole structure. We believe this method simplify computation process since the modes in both sides with the same overlap integral cancels due to the orthogonality product used to establish eigenmodes using complex-conjugated forms of fields in both sides. To execute the proposed 2D problem, the summation of computed eignevectors from formulation of the coupled mode is established in two sides separately. The manuscript go through the theory and some derivations in first section and provides the results and discussion in next segment.

**Theory**

In Figure 1, the discontinuity of metal film in the core of a suggested waveguide is introduced to form an asymmetric waveguide. The perturbation in middle with a metal thin-film supports short range surface plasmons (SRSP) which is surrounded with perfectly matched layers to truncate the field distribution at the edges for remaining the physical domain an ideal window for computing TM modes in "z" axis. The applied perturbation in middle and the boundaries for both sides are shown in Figure 1, and are $\tilde{\varepsilon}_1 = \varepsilon + \Delta\varepsilon_1(x) + \Delta\varepsilon_2(x)$ and $\tilde{\varepsilon}_2 = \varepsilon + \Delta\varepsilon_1(x)$ for left and right hand sides respectively.

Assuming that a shortly time harmonic wave propagates in the z direction as $e^{i(\beta z - \omega t)}$, where $\beta$ is a propagation constant, $\omega$ is the angular frequency. For simplicity, considering short range surface plasmons gives two transverse components in backward and forward directions and ignoring the time dependency of the electric fields we have:

$$e_{tf}(x,z) = A_f(z) \frac{\beta_z H_0}{\omega \varepsilon} e^{i\beta_z z} \cos(\beta_x \cdot x)$$

$$e_{tf}(x,z) = A_b(z) \frac{\beta_z H_0}{\omega \varepsilon} e^{-i\beta_z z} \cos(\beta_x \cdot x)$$

In which, $A_f(z) = a_f e^{-\left(i\frac{n_{eff} \cdot \omega}{c}\right) \cdot z}$ and $A_b(z) = a_b e^{-\left(i\frac{n_{eff} \cdot \omega}{c}\right) \cdot z}$ are forward and backward coefficient indices that indicates the magnitude of modes for each mode. Seemingly, two longitudinal components for both forward and backward modes are as below:

$$e_{zf}(x,z) = A_f(z) \frac{\beta_z H_0}{\omega \varepsilon} e^{i\beta_z z} \sin(\beta_x \cdot x)$$

$$e_{zb}(x,z) = A_b(z) \frac{\beta_z H_0}{\omega \varepsilon} e^{-i\beta_z z} \sin(\beta_x \cdot x)$$

According to the Maxwell's equation for the perturbed waveguides, the magnetic fields for both transvers and longitudinal components follow the equations below:

$$\nabla \times E(x,z) = -i\omega\mu_0 H(x,z)$$
$$\nabla \times H(x,z) = +i\omega\tilde{\varepsilon}(x,z) H(x,z)$$

Thus, the electric and magnetic fields of perturbed waveguides in terms of transverse and longitudinal components are:

$$E_t(x,z) = \sum_{m=1} [a_f(z) + a_b(z)] e_{tm}(x,z)$$

$$H_t(x,z) = \sum_{m=1} [a_f(z) - a_b(z)] h_{tm}(x,z)$$

By plugging these in the Maxwell's equation, the longitudinal components will be:

$$E_z(x,z) = \sum_{m=1} [a_f(z) - a_b(z)] \frac{\varepsilon}{\varepsilon_r} e_{zm}(x,z)$$

$$H_z(x,z) = \sum_{m=1} [a_f(z) + a_b(z)] h_{zm}(x,z)$$

Using transmission matrices, the propagation constant for TM modes in both sides with the defined complex permittivity that is shown in Figure 1 are summarized as below:

$$\beta = \mu\varepsilon\omega^2 = \sqrt{\beta_x^2 + \beta_z^2}$$

$$\beta_x = \frac{m\pi}{d}, (m = 1, 2, 3, \dots)$$

$$\beta_z = \sqrt{\mu\varepsilon\omega^2 + \frac{m\pi}{d}}$$

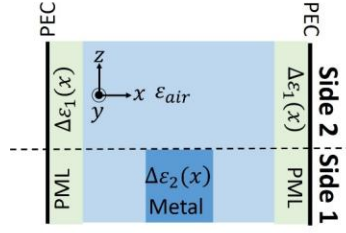

Figure 1. The structure of asymmetric waveguide with perturbing profiles in which the propagation direction is in 'z'axis and there is no 'y' component in the suggested problem; black solid lines in up and down are perfect electric conductor, the gray layers beneath the black solid lines are perfectly matched layers with the permittivity $\Delta\varepsilon_1(x)$ and the dark blue line in side 1 at the core is a thin metal-film with $\Delta\varepsilon_2(x) = 12 + i1.5$

The functionalized complex coupled mode theory for governing the mode amplitude in forward direction is[16]:

$$\frac{\partial}{\partial z}A_{f-n} + i\beta_n A_{f-n} = -i\left(\sum_{m=1}^{10} A_{f-n-m}\frac{\mathcal{X}_{nm}}{\mathcal{N}_m} + \sum_{m=1}^{10} A_{b-n-m}\frac{\mathcal{K}_{nm}}{\mathcal{N}_m}\right)$$

And for the backward direction

$$\frac{\partial}{\partial z}A_{b-n} - i\beta_n A_{b-n} = +i\left(\sum_{m=1}^{10} A_{f-n-m}\frac{\mathcal{X}_{nm}}{\mathcal{N}_m} + \sum_{m=1}^{10} A_{b-n-m}\frac{\mathcal{K}_{nm}}{\mathcal{N}_m}\right)$$

In which the coupling coefficients are given by

$$\mathcal{K}_{mn} = \frac{\omega}{2}\int \Delta\varepsilon(x)([e_{mt}(x,z).e_{nt}(x,z)] - [e_{mz}(x,z).e_{nz}(x,z)])dx$$

$$\mathcal{X}_{mn} = \frac{\omega}{2}\int \Delta\varepsilon(x)([e_{mt}(x,z).e_{nt}(x,z)] + [e_{mz}(x,z).e_{nz}(x,z)])dx$$

Where the perturbing permittivity for both sides are different as it is shown in Figure 1. In addition, the coefficient $\mathcal{N}_m$ is defined as

$$\mathcal{N}_m = \int [e_{mt}(x,z) \times h_{mt}(x,z)].\hat{z}\,dx$$

The expanded complex coupled mode theory can be written in two matrices multiplied to the magnitude of modes as vectors.

$$M_1\vec{V}_1 = -iM_2\vec{V}_2$$

In which $M_1$ contains two terms corresponding to the left side of the equation for forward ($\Omega_j = \frac{\partial}{\partial z} + i\beta_{zj}$) and backward directions $\Omega'_j = \frac{\partial}{\partial z} - i\beta_{zj}$ for jth mode in the waveguide. Here we assume 10 modes of forward and backward as below:

$$M_1 = \begin{bmatrix} \Omega_1 & 0 & 0 & 0 & & 0 & 0 \\ 0 & \ddots & 0 & 0 & & \ddots & \vdots \\ \vdots & \ddots & \Omega_{10} & 0 & & \ddots & 0 \\ 0 & \ddots & 0 & \Omega'_1 & & \ddots & \vdots \\ \vdots & \ddots & 0 & 0 & & \ddots & 0 \\ 0 & 0 & 0 & 0 & & 0 & \Omega'_{10} \end{bmatrix}$$

And $M_2$ coresponds the matrice for the left side of the functionalized complex coupled mode theory that is defined for 10 modes of forward and backward as below

$$M_2 = \begin{bmatrix} +(\mathcal{K}_{1-1}/\mathcal{N}_1) & \cdots & +(\mathcal{K}_{1-10}/\mathcal{N}_1) & +(\mathcal{X}_{1-1}/\mathcal{N}_1) & \cdots & +(\mathcal{X}_{1-10}/\mathcal{N}_1) \\ \vdots & \cdots & \vdots & \vdots & \cdots & \vdots \\ +(\mathcal{K}_{10-1}/\mathcal{N}_{10}) & \cdots & +(\mathcal{K}_{10-10}/\mathcal{N}_{10}) & +(\mathcal{X}_{10-1}/\mathcal{N}_{10}) & \cdots & +(\mathcal{X}_{10-10}/\mathcal{N}_{10}) \\ -(\mathcal{K}_{1-1}/\mathcal{N}_1) & \cdots & -(\mathcal{K}_{1-10}/\mathcal{N}_1) & -(\mathcal{X}_{1-1}/\mathcal{N}_1) & \cdots & -(\mathcal{X}_{1-10}/\mathcal{N}_1) \\ \vdots & \cdots & \vdots & \vdots & \cdots & \vdots \\ -(\mathcal{K}_{10-1}/\mathcal{N}_{10}) & \cdots & -(\mathcal{K}_{10-10}/\mathcal{N}_{10}) & -(\mathcal{X}_{10-1}/\mathcal{N}_{10}) & \cdots & -(\mathcal{X}_{10-10}/\mathcal{N}_{10}) \end{bmatrix}$$

Consequently, the vectors multiplying to the above coefficients representing the magnitude of forward and backward modes are given

$$\vec{V}_1 = \vec{V}_2 = \begin{bmatrix} A_{f-1}(z) \\ \vdots \\ A_{f-10}(z) \\ A_{b-1}(z) \\ \vdots \\ A_{b-10}(z) \end{bmatrix}$$

One of two square matrices ($M_1$) multiplying to the vectors of amplitude is a diagonal matrix that contains the gradient of both forward and backward modes' magnitude and propagation constant of corresponding mode in 'z' direction. Thus, finding eigenvalues of ($M_2$) and consequently Gaussian elimination will give the diagonal matrice of whole eigenvectors in the left side. In this step, the modes amplitude and shape can be determined through the eigenvectors.

## Results and discussion

The purposed structure has a dimension window $80 \times 200$ [$nm$] surrounding with PML and the dimension of rectangular gold film in the middle with the permittivity of $-12 + i1.5$ is $40 \times 50$ [$nm$] truncated in the middle of the waveguide. The air around the metal film makes an asymmetric composite. Using finite difference time domain, the effective index for 10 TM modes are shown in Figure 2.

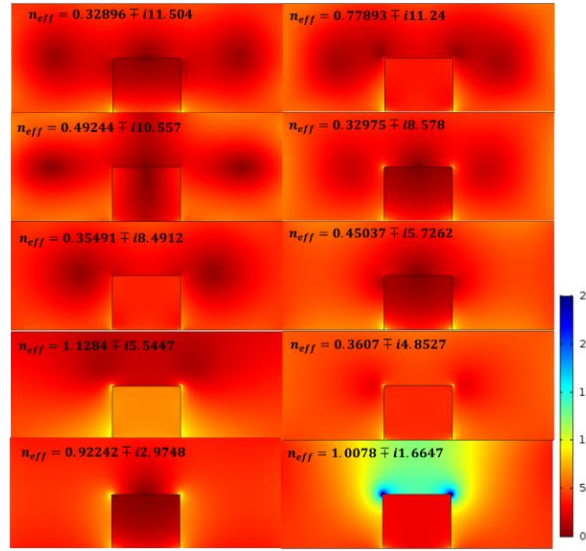

Figure 2. FDTD computation results for 10 TM modes in a thin-film surrounded with air and PML at the boundries.

According to the simulation results, the electric field distribution for all modes jump near the edge of the thin film and gives rise to the generation of modes.

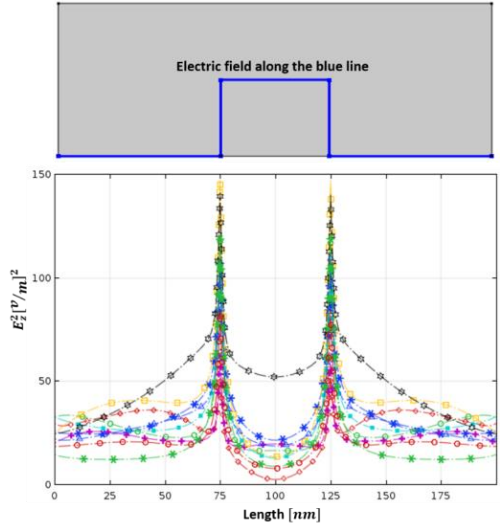

Figure 3. Field distribution of modes along the shown blue line of the structure above the graph

However, the purpose of this study is investigating analytically and numerically the dynamics of the TM modes for a lossy thin film near the PML layers. The applied perfectly matched layer profile follows the equation below to connect permittivity of layer to the conductivity.

$$\sigma = \sigma_{max}(\frac{\rho}{T_{PML}})^m$$

In which, m=2 and $\rho$ is the distance from measuring point and $T_{PML}$ is the thickness of layer. By considering $\sigma = -i\omega\varepsilon_0\varepsilon_r$ the conductivity of the PML, the relation can be re-written as below

$$-i\omega\varepsilon_0\varepsilon_r = (\frac{x}{2E-6})^2$$

Therefore, the perturbing layer possesses the permittivity of $\Delta\varepsilon_1 = (\frac{1}{\omega*4E-12})x^2$, so, for the left and right hand side the permittivity follows the function of 'x'. For the left hand side we can assume $\tilde{\varepsilon}_1 = -12 + i1.5 + (\frac{1}{\omega*4E-12})x^2$ and for the right hand side $\tilde{\varepsilon}_2 = (\frac{1}{\omega*4E-12})x^2$.

For the analytical calculation of the coefficient integrals, the constant term is $a_f a_b e^{-i(\frac{n_{eff}}{c}).z}\left(\frac{\beta_z H_0}{\omega c}\right)^2$ for both sides and the solutions for ten modes in LHS

$$\mathcal{K}_{mn} = (-2x^2+(\beta_{xm}+\beta_{xn})^2)sin((\beta_{xm}+\beta_{xn})x) + 2(\beta_{xm}+\beta_{xn}).x.cos((\beta_{xm}+\beta_{xn})x) + sin(x.\beta_{xm}+\beta_{xn})$$

And for the RHS

$$\mathcal{K}_{mn} = \frac{1}{(\beta_{xm}+\beta_{xn})^3}.(-2+x^2(\beta_{xm}+\beta_{xn})^2)sin((\beta_{xm}+\beta_{xn})x) + 2(\beta_{xm}+\beta_{xn}).x.cos((\beta_{xm}+\beta_{xn})x)$$

, likewise LHS

$$\mathcal{X}_{mn} = (-2x^2+(\beta_{xm}-\beta_{xn})^2)sin((\beta_{xm}-\beta_{xn})x) + 2(\beta_{xm}-\beta_{xn}).x.cos((\beta_{xm}-\beta_{xn})x) + sin(x.\beta_{xm}-\beta_{xn})$$

And for RHS

$$\mathcal{X}_{mn} = \frac{1}{(\beta_{xm}-\beta_{xn})^3}.(-2+x^2(\beta_{xm}-\beta_{xn})^2)sin((\beta_{xm}-\beta_{xn})x) + 2(\beta_{xm}-\beta_{xn}).x.cos((\beta_{xm}-\beta_{xn})x)$$

Finally, the analytic calculation for the norm of the m$^{th}$ mode is $\mathcal{N}_m = \int[e_{mt}(x,z) \times h_{mt}(x,z)].\hat{z}\,dx$ that is a cross product of fields in the propegation direction

$$\begin{vmatrix} \hat{x} & \hat{y} & \hat{z} \\ 0 & e_t & 0 \\ h_t & 0 & 0 \end{vmatrix} = A_f(z)B_f(z)\frac{\beta_z H_0^2}{\omega \varepsilon}e^{(i2\beta_z).z}\cos^2(\beta_x . x)\hat{z}$$

And likewise for backward modes

$$\begin{vmatrix} \hat{x} & \hat{y} & \hat{z} \\ 0 & e_t & 0 \\ h_t & 0 & 0 \end{vmatrix} = A_b(z)B_b(z)\frac{\beta_z H_0^2}{\omega \varepsilon}e^{-(i2\beta_z).z}\sin^2(\beta_x . x)\hat{z}$$

After assigning the values and finding the coefficient of the C-CMT formula, the eigenvectors corresponding to the each modes gives the electric field distribution in one dimension as we assumed already. Figure 4 shows the numerical results for 10 TM modes through the length of the structure in side 1. The Figure 4, shows computed electric field distribution for all modes after finding the magnitude of both forward and backward eigenvectors in a half symmetry of the structure starting from the boundary that the field is truncated to the core of waveguide.

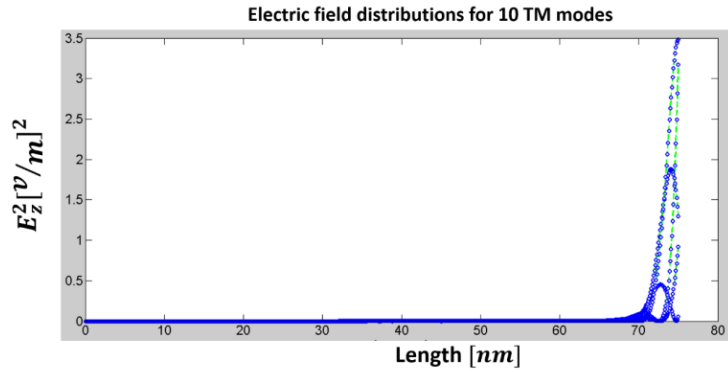

Figure 4. Numerical computed Electric field distribution squared for 10 TM modes

The same approach for the air side of the structure gives the field distribution for 10 TM modes.

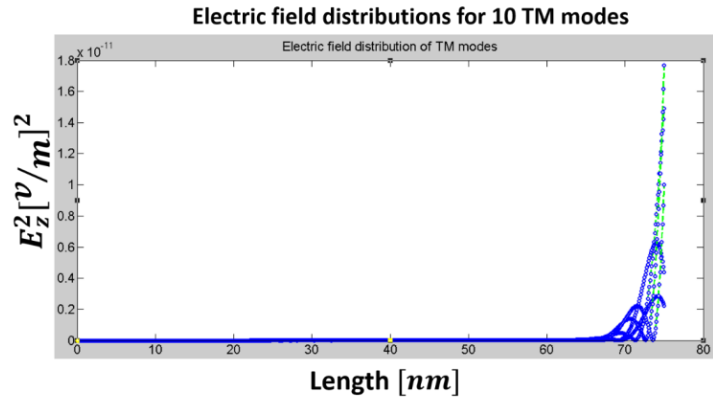

Figure 5. Numerical computed Electric field distribution squared for 10 TM modes in the airside of the composite

Using Mode Matching (MM) we can find the transmission and reflection coefficient in the suggested 2D problem. For that, the summation of all transmitted in one side for both field is as below

$$\vec{E}_{t,1} = \sum_m (A_{f-m,1} + A_{b-m,1})\vec{E}_{m,1}$$

And for magnitude fields of transmitted modes

$$\vec{H}_{t,1} = \sum_m (A_{f-m,1} - A_{b-m,1})\vec{H}_{m,1}$$

In addition, we can use the same equations for the modes in side 2 as below

$$\vec{E}_{t,2} = \sum_n (A_{f-n,2} + A_{b-n,2})\vec{E}_{n,2}(x,z)$$

And for the magnetic field

$$\vec{H}_{t,2} = \sum_n (A_{f-n,2} - A_{b-n,2})\vec{H}_{n,2}(x,z)$$

By considering the continuity, conditions $E_{t,1} = E_{t,2}$ and $H_{t,1} = H_{t,2}$ the satisfying conjugated equation govern the mode matching method. In other words, conjugated orthogonality of a mode in both sides get zero value and cancel out in both sides of the summation equation. Thus, the result for transmission and reflection coefficient are as below:

$$\sum_{n,m} \int (\vec{E}_{n,1}(x,z) \times \vec{H}^*_{m,2}(x,z))dx = 0$$

We already computed the modes amplitude numerically and assigning those in the fields we have trnsmision coefficient of

$$E^i_x(x,z) = A_{f-n} \cos(\beta_{x-n} \cdot x) \, e^{+i(\beta_z \cdot z)} \, \hat{x}$$

$$H^i_y(x,z) = \frac{\kappa_0 \, \xi_0}{\beta_{y-n}} A_{f-n} \sin(\beta_{x-n} \cdot x) \, e^{+i(\beta_z \cdot z)} \, \hat{y}$$

Then, we have reflection coefficient of

$$E^r_x(x,z) = A_{b-n} \cos(\beta_{x-n} \cdot x) \, e^{+i(\beta_z \cdot z)} \, \hat{x}$$

$$H^r_y(x,z) = \frac{\kappa_0 \, \xi_0}{\beta_{y-n}} A_{b-n} \sin(\beta_{x-n} \cdot x) \, e^{+i(\beta_z \cdot z)} \, \hat{y}$$

In **Figure 6**, the results for the transmission and reflection coefficient of modes are plotted according to the summation of forward and backward modes in two sides.

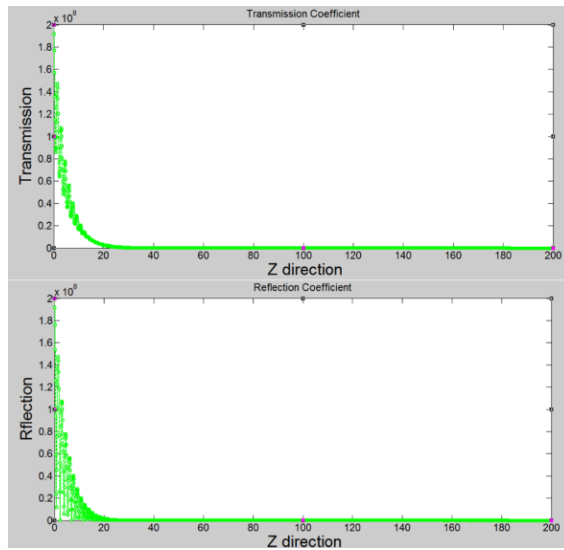

Figure 6. Transmission and reflection coefficient profile along the z direction

## Conclusion

We investigated complex coupled mode theory for an asymmetric planar waveguide containing metal thin-film as a temporal resonator at the core surrounded with air and perfectly matched layers. Since the coupled mode theory is a cumbersome method to solve structures with lossy materials in which radiation fields or lossy materials that eventuate in establishment of the leaky modes due to the fact that the continuum of radiation is not orthogonal and normalizable in real domain. The presented 2D problem in this study provides significance of such method implementing on an asymmetric optical elements that are in high interesting scope of integrated optical resonators and fiber Bragg gratings.

## References

[1]     Z. Ruan, S. Fan, Temporal coupled-mode theory for fano resonance in light scattering by a single obstacle, J. Phys. Chem. C. 114 (2010) 7324–7329. doi:10.1021/jp9089722.

[2]     L. Verslegers, Z. Yu, Z. Ruan, P.B. Catrysse, S. Fan, From electromagnetically induced transparency to superscattering with a single structure: A coupled-mode theory for doubly resonant structures, Phys. Rev. Lett. 108 (2012). doi:10.1103/PhysRevLett.108.083902.

[3]     H.A. Haus, W. Huang, Coupled-Mode Theory, Proc. IEEE. 79 (1991) 1505–1518. doi:10.1109/5.104225.

[4]     S. Fan, W. Suh, J.D. Joannopoulos, Temporal coupled-mode theory for the Fano resonance in optical resonators, J. Opt. Soc. Am. A. 20 (2003) 569. doi:10.1364/josaa.20.000569.

[5]     Y.-C. Lu, W.-P. Huang, S.-S. Jian, Full Vector Complex Coupled Mode Theory for Tilted Fiber Gratings, Opt. Express. 18 (2010) 713. doi:10.1364/oe.18.000713.

[6]     Y. Jeong, B. Lee, Nonlinear property analysis of long-period fiber gratings using discretized coupled-mode theory, IEEE J. Quantum Electron. 35 (1999) 1284–1292. doi:10.1109/3.784588.

[7]     A. Karalis, J.D. Joannopoulos, Temporal coupled-mode theory model for resonant near-field thermophotovoltaics, Appl. Phys. Lett. 107 (2015). doi:10.1063/1.4932520.

[8]     D.N. Maksimov, A.F. Sadreev, A.A. Lyapina, A.S. Pilipchuk, Coupled mode theory for acoustic resonators, Wave Motion. 56 (2015) 52–66. doi:10.1016/j.wavemoti.2015.02.003.

[9]     S. Khan, H.E. Türeci, Non-Hermitian coupled-mode theory for incoherently pumped exciton-polariton condensates, Phys. Rev. A. 94 (2016). doi:10.1103/PhysRevA.94.053856.

[10]    B. Wu, B. Wu, J. Xu, J. Xiao, Y. Chen, Coupled mode theory in non-Hermitian optical cavities, Opt. Express. 24 (2016) 16566. doi:10.1364/oe.24.016566.

[11]    J. Xu, Y. Chen, General coupled mode theory in non-Hermitian waveguides, Opt. Express. 23 (2015) 22619. doi:10.1364/oe.23.022619.

[12]    M. Mlejnek, I. Roudas, J.D. Downie, N. Kaliteevskiy, K. Koreshkov, Coupled-mode theory of multipath interference in quasi-single mode fibers, IEEE Photonics J. 7 (2015). doi:10.1109/JPHOT.2014.2387260.

[13]    S.Y. Elnaggar, R.J. Tervo, S.M. Mattar, Energy Coupled Mode Theory for Electromagnetic Resonators, IEEE Trans. Microw. Theory Tech. 63 (2015) 2115–2123. doi:10.1109/TMTT.2015.2434377.

[14]    T. Christopoulos, O. Tsilipakos, N. Grivas, E.E. Kriezis, Coupled-mode-theory framework for nonlinear resonators comprising graphene, Phys. Rev. E. 94 (2016). doi:10.1103/PhysRevE.94.062219.

[15]    T.E.-J. of lightwave technology,  undefined 1997, Fiber grating spectra, Ieeexplore.Ieee.Org. (n.d.). https://ieeexplore.ieee.org/abstract/document/618322/ (accessed December 5, 2019).

[16] W.-P. Huang, J. Mu, Complex coupled-mode theory for optical waveguides, Opt. Express. 17 (2009) 19134. doi:10.1364/oe.17.019134.

[17] T. DeWolf, R. Gordon, Complex coupled mode theory electromagnetic mode solver, Opt. Express. 25 (2017) 28337. doi:10.1364/oe.25.028337.

[18] W.-P. Huang, J. Mu, Simulation of radiation coupling by complex coupled-mode theory, in: Optoelectron. Mater. Devices V, OSA, Washington, D.C., 2010: p. 798705. doi:10.1364/ACP.2010.798705.

[19] T. Erdogan, Cladding-mode resonances in short- and long-period fiber grating filters, J. Opt. Soc. Am. A. 14 (1997) 1760. doi:10.1364/josaa.14.001760.

[20] T. Erdogan, Fiber grating spectra, J. Light. Technol. 15 (1997) 1277–1294. doi:10.1109/50.618322.

[21] C. Lu, Y. Cui, Fiber Bragg Grating Spectra in Multimode Optical Fibers, J. Light. Technol. Vol. 24, Issue 1, Pp. 598-. 24 (2006) 598.

[22] R. Sammut, A.W. Snyder, Leaky modes on a dielectric waveguide: orthogonality and excitation, Appl. Opt. 15 (1976) 1040. doi:10.1364/ao.15.001040.

[23] A. Snyder, J. Love, Optical waveguide theory, 2012. https://books.google.com/books?hl=en&lr=&id=DCXVBwAAQBAJ&oi=fnd&pg=PA3&dq=Optical+waveguide+Theory&ots=xaupi-Numj&sig=GodUiXEVihPc2T6hkSGDGpjoKTA (accessed December 6, 2019).

[24] A. Hardy, W. Streifer, Coupled Mode Theory of Parallel Waveguides, J. Light. Technol. 3 (1985) 1135–1146. doi:10.1109/JLT.1985.1074291.